\documentclass[prl,twocolumn,floatfix,showpacs]{revtex4}
\usepackage{graphicx,natbib}
\begin{document}


\title{Off Equilibrium Study of the Fluctuation-Dissipation Relation in the Easy-Axis Heisenberg
Antiferromagnet on the Kagome Lattice}


\author{S. Bekhechi and B.W. Southern}
\affiliation{Department of Physics and Astronomy\\ University of Manitoba \\ Winnipeg Manitoba \\ Canada R3T 2N2}


\date{\today}

\begin{abstract}
Violation of the fluctuation-dissipation theorem (FDT) in a frustrated Heisenberg model on the Kagome
lattice is investigated using Monte Carlo simulations. The model exhibits glassy behaviour at low
temperatures accompanied by very slow dynamics. Both the spin-spin autocorrelation function and the response
to an external magnetic field are studied. Clear evidence of a constant value of the fluctuation
dissipation ratio and long range memory effects are observed for the first time in this model. The breakdown of the FDT in the glassy phase follows the
predictions of the mean field theory for spin glasses with one-step replica symmetry breaking.
\end{abstract}


\maketitle


Glassy systems are characterized by long relaxation times which increase at low temperatures and eventually exceed
experimental observation times. At higher temperatures, the systems are ergodic and true equilibrium can be reached. However,
below the glass transition temperature, $T_g$, the systems can become trapped in metastable states and  often display strong
 non-equilibrium
effects such as {\it aging}. The term {\it aging} refers to the fact that both static and dynamic properties depend on the time interval after which
 the system has been quenched into some
 non-equilibrium state.   Structural glasses \cite{n6b} and spin glasses \cite{n6a}  exhibit these features and are
 essentially out of equilibrium on experimental time scales. In the case of spin systems, this behaviour
 is usually found in sytems with  some type of quenched disorder. The disorder often imposes severe restrictions on the spin rearrangements and the
evolution towards equilibrium requires many degrees of freedom to act cooperatively. The complexity of the phase space can lead to
the violation of the fluctuation dissipation theorem (FDT) below a characteristic temperature.

In the last decade off equilibrium approaches have been used quite succesfully  to describe systems which 
show a very slow dynamics \cite{rev}.  A typical
 example is the zero-field cooling \cite{n6c} experiment, in which a sample is cooled
 in zero magnetic field to a low temperature at time $t=0$. After a waiting time $t_w$
 a small magnetic field is applied and subsequently the time evolution of the
 magnetization is recorded. It is often observed that the relaxation of the magnetization becomes slower as
 the waiting time $t_w$ is increased. 
 The system never reaches
 thermodynamic equilibrium and, consequently, time translational invariance (TTI) and the fluctuation
 dissipation theorem  are no longer valid.  However, the system is in a quasi-equilibrium state for time scales
which are much longer than microscopic time scales but shorter than the relaxtional time scale of the slowly relaxing degrees of
freedom. In this glassy regime a  generalized form of the FDT has been proposed
 \cite{n7a,n7b},
 \begin{equation}
R(t+t_w,t_w)=\frac{X(t+t_w,t_w)}{T} \frac{\partial C(t+t_w,t_w)}{\partial t_w}
\end{equation}
where $C(t+t_w,t_w)$ is a double time correlation function and $R(t+t_w, t_w)$ with $t > 0$ is the
associated conjugate response function.  In this relation, $T$ is the heat bath temperature and the function $X(t+t_w,t_w)$ is called the
fluctuation dissipation ratio (FDR)
and is a measure of the departure from equilibrium. In the equilibrium regime, $X$ is equal to unity and the usual FDT is
recovered. In the out of equilibrium regime, $X$ is generally less than unity.  For large values of $t$ and
$t_w$, it has been hypothesized \cite{n7c} that $X$ becomes a function of time only through the correlation function $C$  and a Quasi-FDT (QFDT)  relation is obtained,
 \begin{equation}
R(t+t_w,t_w)=\frac{X(C)}{T} \frac{\partial C(t+t_w,t_w)}{\partial t_w}
\end{equation}

An important property of this function $X(C)$ is that it provides indirect information on the
structure of the phase space.  The ratio $T_{eff}(C)= \frac{T}{X(C)}$ has been interpreted \cite{n7d} as an effective
temperature which is generally larger than the heat bath temperature $T$. This effective temperature has been used to classify glassy systems  into three main groups: 
 in models with a single pure
state such as a pure ferromagnet \cite{n8a} and random Ising systems \cite{n8b}, $T_{eff}(C)$ is infinite since $X=0$ ; 
in models of structural 
glasses \cite{n8aa,n8bb}, Lennard-Jones glasses \cite{n8c,n8d} and $p$-spin models in $d=3$ \cite{n8e}, $T_{eff}(C) > T$ is finite; 
in systems with full replica symmetry breaking as in several  finite dimensional spin glasses
\cite{n8f,n8g,n8h} and mean field (MF) models for spin glasses \cite{n8i}, $T_{eff}(C)$ 
is a nontrivial function of $C$.

Geometrically frustrated antiferromagnets can also exhibit glassy behaviour and other novel kinds of
low temperature magnetic states  \cite{n1a,n1b,n1c} which are quite different from those observed in conventional
magnets. 
Recently we have performed a  numerical study of the two dimensional easy-axis Heisenberg 
antiferromagnet on the Kagome lattice  \cite{n5e} by computing its static and dynamic properties. The magnetization indicates a finite $T_c$ but Monte Carlo
snapshots of the individual spins below this temperature do not indicate any long ranged spatial order. Rather, the individual spins
appear to be in a frozen state similar to a glass. The three spins on each elementary triangle form a distorted $120^0$ planar
state with a net magnetization in the $z$-direction but there is no sublattice order.
We  extracted the critical temperature and the critical exponents associated with the
magnetization, the susceptibility and the correlation length.  We  also studied the two-time spin-spin correlation function at high and low temperatures. We have found that non-exponential relaxation sets in at a temperature $T^* > T_c$. The relaxation time increases according  to a power law and diverges at a
 temperature $T_g < T_c$ where a transition to a glassy phase is located. 
Below $T_g$, 
  we have found clear
 evidence for the presence of aging effects in the autocorrelation function from
 off-equilibrium dynamics.  The aging effects  obey the
same scaling laws that are observed in spin glasses  and polymers. 

In this letter we report the first observation of the violation of the FDT in this low dimensional geometrically 
frustrated system without disorder. 
We determine both  the FDR and the effective temperature $T_{eff}(C)$.
Our results for $X(C)$ indicate that the system behaves like a generalized mean field model of glasses in which there is  a one step replica symmetry breaking. In the aging regime, $X(C)$ has an approximate linear 
dependence on the temperature.  

 The Hamiltonian describing the model is,
\begin{equation}
H = J\sum_{i<j} (S_{i}^{x} S_{j}^{x} + S_{i}^{y} S_{j}^{y}+ A S_{i}^{z}
S_{j}^{z})
\end{equation}
where $(S_{i}^{\alpha},\alpha=x,y,z)$ represents a classical three component
spin of unit magnitude located at each site $i$ of a Kagome lattice. The number of
sites $N=3 L^2$ where $L$ is the linear size of the lattice and the
exchange interactions are restricted to
nearest-neighbour pairs of sites. The parameter $A$ describes the strength of the exchange anisotropy. 
 In the following we restrict
our consideration to the value $A=2$ which corresponds to the easy axis case.

In order to study the non-equilibrium properties of this easy axis frustrated system, we  calculate the spin-spin
 autocorrelation function \footnote{This is the same autocorrelation function studied in reference \cite{n5e} but the
notation for the arguments is slightly different.}
  \begin{equation}
C(t+t_w,t_w)=\frac{1}{N} \sum_{i} <S_{i}^{z}(t+t_w) S_{i}^{z}(t_w)>
\end{equation}
 where $<...>$ means an average over thermal histories and $t_w$ is the waiting
 time measured from some quenching time $t_o=0$. A second quantity of interest is the associated
 response function to an external magnetic field $h_i(t)$ applied at $t_w$,
\begin{equation}
M(t+t_w,t_w)=\int_{t_w}^{t+t_w}R(t+t_w,s)h(s)ds.
\end{equation}
 The off equilibrium susceptibility $\chi(t+t_w,t_w)$ can obtained by dividing by the field magnitude and the QFDT relation (2) allows
 us to write  for long times $(t, t_w  \rightarrow \infty)$
   \begin{equation}
	T \chi(t+t_w,t_w)=\frac{T}{h} M(t+t_w,t_w)=\int_{C(t+t_w,t_w)}^{C_0} X(C) dC. 
\end{equation}

If the FDT is satisfied ($X(C)=1$), then we obtain the  linear relation $T \chi(t+t_w,t_w)=C_0-C(t+t_w,t_w)$.
 A departure from this straight line in a $T\chi$ vs $C$ parametric plot
indicates a violation of the FDT and yields information about the function $X(C)$.
In Ising systems, the upper limit $C_0$ in (6) is unity but in this Heisenberg system the value depends on both $T$ and the value
of the easy axis anisotropy $A$. At $T=0$ it is given   by $C_0=\frac{(A+1)^2+2 A^2}{3 (A+1)^2}$ but must be determined numerically at finite $T$.

\begin{figure}[b]
\centering
\includegraphics[height=65mm,width=85mm]{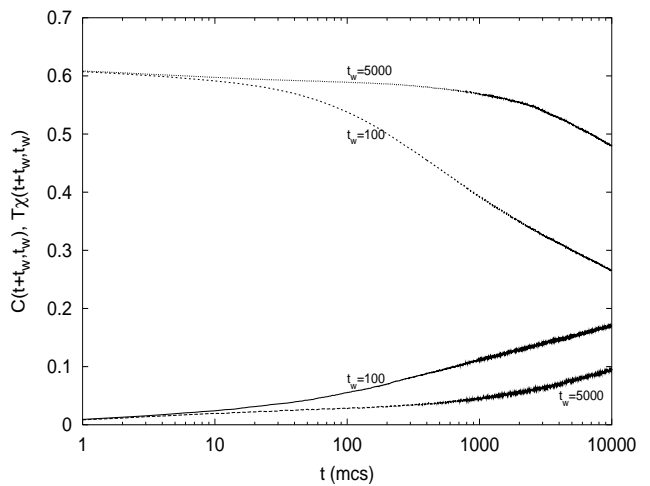}
\caption{Correlation function, $C(t+t_w,t_w)$ ( upper curves), and the integrated response function,
$T\chi(t+t_w,t_w)$, (lower curves) at $T=0.04$ and $h^z=0.01$ for  waiting times $t_w=100, 5000$}
\end{figure}

For each Monte Carlo run, the system is initialized in a random initial configuration
 corresponding to a quenching from infinite temperature to the temperature $T$. We then
allow the system to evolve for a time $t_w$ and we then make a second copy of the system. Using the
original copy  we compute
 $C(t+t_w,t_w)$ as a function of the observation time $t>0$, for different values of $t_w$ and $T$. 
In the second copy we apply  a  perturbation in the form of random small magnetic field $h(i)=h^z
 \epsilon_i$  in order to avoid favoring one of the different phases \cite{n8d}.
 The $\epsilon_i$ are taken from a bimodal distribution $(\epsilon_{i}=\pm 1)$ and the strength of the
 field $h^z$ is taken to be small (typically between $h^z$=0.002 to 0.01) to ensure linear response. We 
  compute the staggered magnetization in this second copy
 \begin{equation}
	M(t+t_w,t_w)=\frac{1}{N} \sum_{i} \overline{<S_{i}^{z}(t+t_w) \epsilon_i>}, 
\end{equation}
whose conjugate field has magnitude $h^z$ and where the overline means an average over the random variables
$\epsilon_{i}$.

\begin{figure}[t]
\centering
\includegraphics[height=65mm,width=85mm]{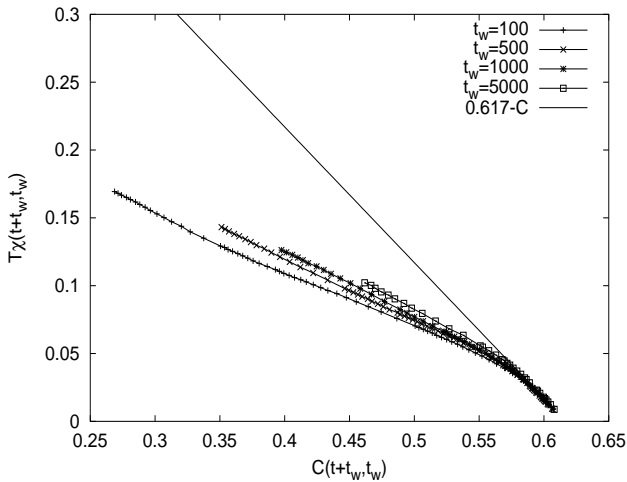}
\caption{The integrated response vs the correlation function at
$T=0.04$ for waiting times $t_w=100, 500, 1000, 5000$ and field strength $h^z=0.01$ with $L=60$. 
The straight line corresponds to $T\chi = 0.617 - C$.}
\end{figure}

Typical data for the integrated response and the correlation function are shown in Fig. 1 
  where both
  functions exhibit a stationary part for at short times $t << t_w$ and an aging part for $t >> t_w$.
 In Fig.2 , a parametric plot of the integrated response versus autocorrelation function is shown for
  $T=0.04$ obtained with a perturbing field magnitude $h^z=0.01$, for waiting times $t_w=100, 500, 1000$ and
 $5000$ Monte Carlo steps (mcs). It is clearly seen that for short times ($t<<t_w$)  TTI and FDT hold 
 and the straight line has a slope of $-1$. For larger times ($t >> t_w$)  a departure from 
 the FDT line is observed indicating that the system has fallen  out of equilibrium.
 A closer look at the data in the aging regime shows that , while 
 the spin-spin autocorrelation function, $C(t+t_w,t_w)$ is decaying to zero as $t \rightarrow 
 \infty$, the associated
 response function $M(t+t_w,t_w)$  keeps growing 
  for all  waiting times $t_w$ and its value
 depends on $t_w$.  These features are reflected in a nontrivial function $X(C)$ . 
Similar curves have been obtained in previous studies of several
 glassy systems \cite{n8bb,n8c,n8d,n8e} such as $p$-spin models and
structural glasses.
  
\begin{figure}[t]
\centering
\includegraphics[height=65mm,width=85mm]{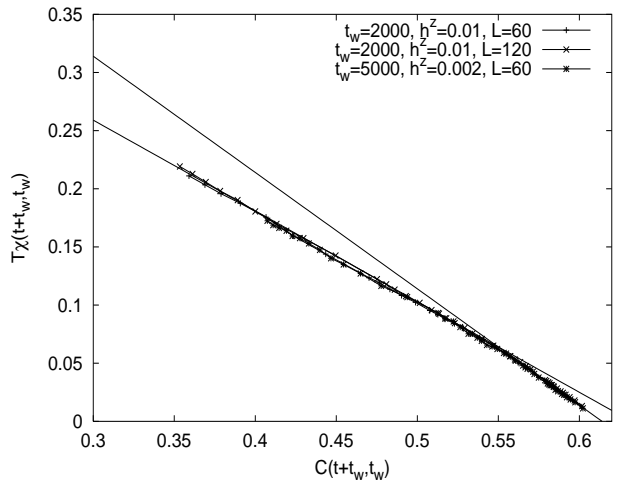}
\caption{The integrated response vs the correlation function at
$T=0.055$ for various sizes $L$, waiting times $t_w$ and field strengths $h^z$. The value of $C_0=.614$ and 
the two solid lines interesect at $C=q_{EA}=.55$ with
 slopes $-1$ 
and $-0.78$ .}
\end{figure}

   The curves in Fig. 2 
show a strong dependence on $t_w$, indicating that we are far from the 
asymptotic regime. This dependence  on $t_w$    could possibly be
 explained by the fact that we are using waiting times that are too small \cite{n8d}.
Indeed, by performing simulations at a slightly  higher temperature $T=0.055$
 we  see  in Fig.3 that the curves  clearly indicate:  (i) two well separated time  scales,
 (ii) no dependence of the FDR on $t_w$ in the aging regime, and (iii) no finite size effects. 
The presence of two time scales indicates  that the system falls
out of equilibrium at the point where the two straight lines intersect. If we identify this point with the
$t_w \rightarrow \infty$ limit of the two-time autocorrelation function, then the FDR $X(C)$ can be described  \cite{n8e} 
in the two regions as follows
 \begin{equation}
	T\chi(C) = \left\{ \begin{array}{ll} C_0 - C, & \mbox{$C \ge q_{EA}$} \\
					     X[q_{EA}-C]+[C_0-q_{EA}], & \mbox{$C \le q_{EA}$}
					     \end{array}
	\right. 
 \end{equation}
where $q_{EA}$  is the value of $C$ at the crossing point and plays a role similar to that of  the Edwards-Anderson order parameter 
in the mean field theory of spin glasses \cite{n10}.  Fitting the data in the out of equilibrium regime to the
  straight line of eqn(8), will give us both the FDR $X(C)$ and $q_{EA}$. 

\begin{figure}[t]
\centering
\includegraphics[height=65mm,width=85mm]{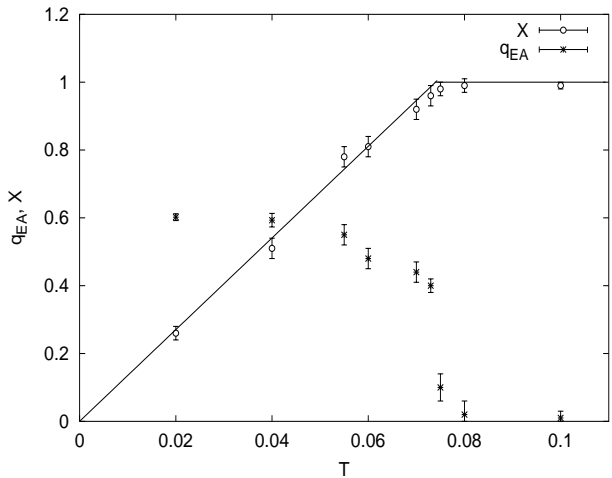}
\caption{The quantities X and $q_{EA}$ for $t_w=2000$ as functions of the temperature.
The straight line is the prediction of the approximation $X(T)=T/T_{eff} $.}
\end{figure}

  The  values of $X(C)$ and $q_{EA}$ obtained from this procedure are displayed in Fig. 4  as a function of $T$ using the data for $t_w=2000$ and $L=60$. We can see
  that at high temperatures $X$ is equal to $1$ and the usual FDT holds, whereas
  for $T<T_{eff}$ our results are well approximated by the straight line 
  $X(T)=T/T_{eff}$ with $T_{eff}=0.074 \pm 0.001$ . As the temperature increases the intersection point  $q_{EA}$ decreases 
  to zero with a sharp drop  at $T_{eff}$ for this lattice size $L=60$.  Our previous results \cite{n5e}  
 for this model indicated a glass transition at a slightly lower 
temperature  $T_g= 0.071 \pm 0.002$ where the relaxation time  diverged. One would expect that $T_{eff} > T_g$ for
systems with finite values of $L$ and $t_w$ and that it will approach $T_g$ for larger system sizes and longer waiting times. An extension of the present work to larger values of both $L$ and $t_w$
will also help to determine whether  $q_{EA}$ approaches zero continuously or discontinuously at the glass transition.

  In summary, we have shown that the fluctuation-dissipation ratio can be computed in this disorder free
 Kagome easy axis antiferromagnet. Several nontrivial features predicted by the  mean field theory
 spin glasses are also present in this model. Namely, long-ranged memory effects and a constant FDR
 $X(C)$. These features correspond to a phase space structure similar to that of spin systems undergoing a one
 step replica symmetry breaking. The temperature dependence of the FDR $X(T)$ varies linearly  with temperature $X(T)=T/T_{eff}$ as observed in binary mixtures of soft spheres \cite{n8c} and Lennard-Jones glasses \cite{n101}.
 All of our results are consistent with the existence of a glassy state at low temperatures in this geometrically
frustrated model.  The ideas developed for mean field descriptions of glassy systems seem to have an application far
beyond the original models. Further experimental studies of these low dimensional frustrated magnets could provide
some insight into the behaviour of glassy materials.

 \begin{acknowledgments}

This work was supported by the Natural Sciences and Research Council of Canada 
and the High Performance Computing facility at the University of Manitoba. 

\end{acknowledgments}

\bibliography{smaine2}

\end{document}